# Coupling tunable D-band directional coupler for millimeter-wave applications

Weiye Xu, Handong Xu and Fukun Liu

A coupling tunable D-band directional coupler is designed based on a novel coupling grid structure proposed in this letter. The designed directional coupler has excellent performance with ultra-wideband. The coupling can be tuned from -28.2 dB to -33.2 dB at 140 GHz by changing the angle of the coupling grid, and the dynamic range of the coupling is about 5 dB. The return loss is smaller than -15 dB in the whole D-band from 110 GHz to 170 GHz. A 3-dB coupler use the similar coupling structure is also designed. The coupling is 3.3144 dB at the center frequency of 140 GHz.

*Introduction:* Microwave directional coupler is a key microwave transmission device in many fields such as wireless communication, radar, and nuclear fusion plasma heating. Directional couplers can be implemented with a variety of transmission lines, such as coaxial line, rectangular waveguide, circular waveguide, stripline, microstrip line, and so on. Among them, the rectangular waveguide is the most widely used transmission line suitable for transmitting waves from several GHz to several hundreds of GHz.

At present, the directional couplers suitable for waves from several GHz to several hundreds of GHz are mostly based on hole coupling [1, 2], waveguide T-Junction [3], etc. The bandwidth of the conventional design is usually small and the coupling is mostly fixed, which is inconvenient for ultra-wideband or large dynamic range applications.

In this letter, a novel ultra-wideband microwave directional coupler is proposed to remedy the defects of the conventional technology. The designed directional coupler has small return loss, good directivity, and the coupling can be tuned and the tuning range is large.

*Directional coupler design:* The designed D-band directional coupler is shown in Fig. 1. The coupler includes three parts: the WR-6 rectangular waveguides, the matching structures, and the coupling grid structure that consists of five metal strips. Where the matching structure is a miter bend, which is to ensure the wide side of the waveguide stays constant. The miter bends are only used between Port 1 and Port2, and between Port 3 and Port 4. The metal strips are made of copper-plated material. The five metal strips are distributed symmetrically and equidistantly. The distance between the centers of two adjacent metal strips is s=0.396 mm. All metal strips have the same thickness, which is 0.02 mm. If the five metal strips are numbered from left to right as No. 1 to No. 5. No. 1 and No. 5 have the same width, which is d1=0.1328 mm. No. 2, No. 3, and No. 4 also have the same width, which is d2=0.0913 mm. The coupling can be controlled by adjusting the angle of the coupling grid. The angle of the coupling grid shown in Fig. 1 is 90 ° or -90 °. When the coupling grid rotates counterclockwise, the angle changes from -90 ° to 90 °.

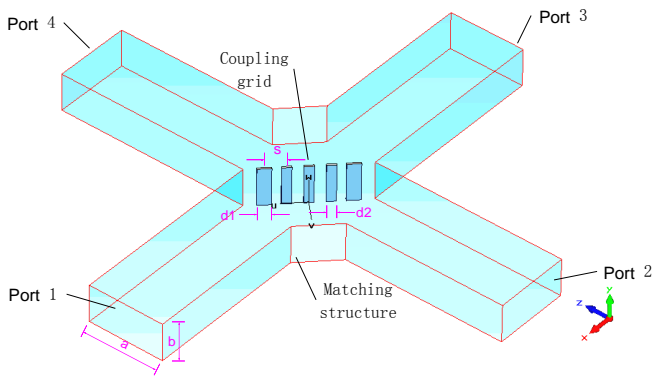

**Fig. 1** *Structure of the designed directional coupler. Where a=1.651 mm, b=0.8255 mm, s=0.396 mm, d1=0.1328 mm, d2=0.0913 mm.*

The simulation results of the designed directional coupler are shown in Fig. 2. The return loss is smaller than -15 dB in the whole D-band from 110 GHz to 170 GHz. The insertion loss S21 is greater than -0.55 dB in D-band and greater than -0.15 dB at the center frequency of 140 GHz. The coupling S31 changes from -28.2 dB to -33.2 dB at the center frequency with the angle of the coupling grid. The dynamic range of the coupling is about 5 dB. The directivity S31-S41 is greater than 9 dB in D-band and the directivity is greater than 16 dB at the center frequency of 140 GHz.

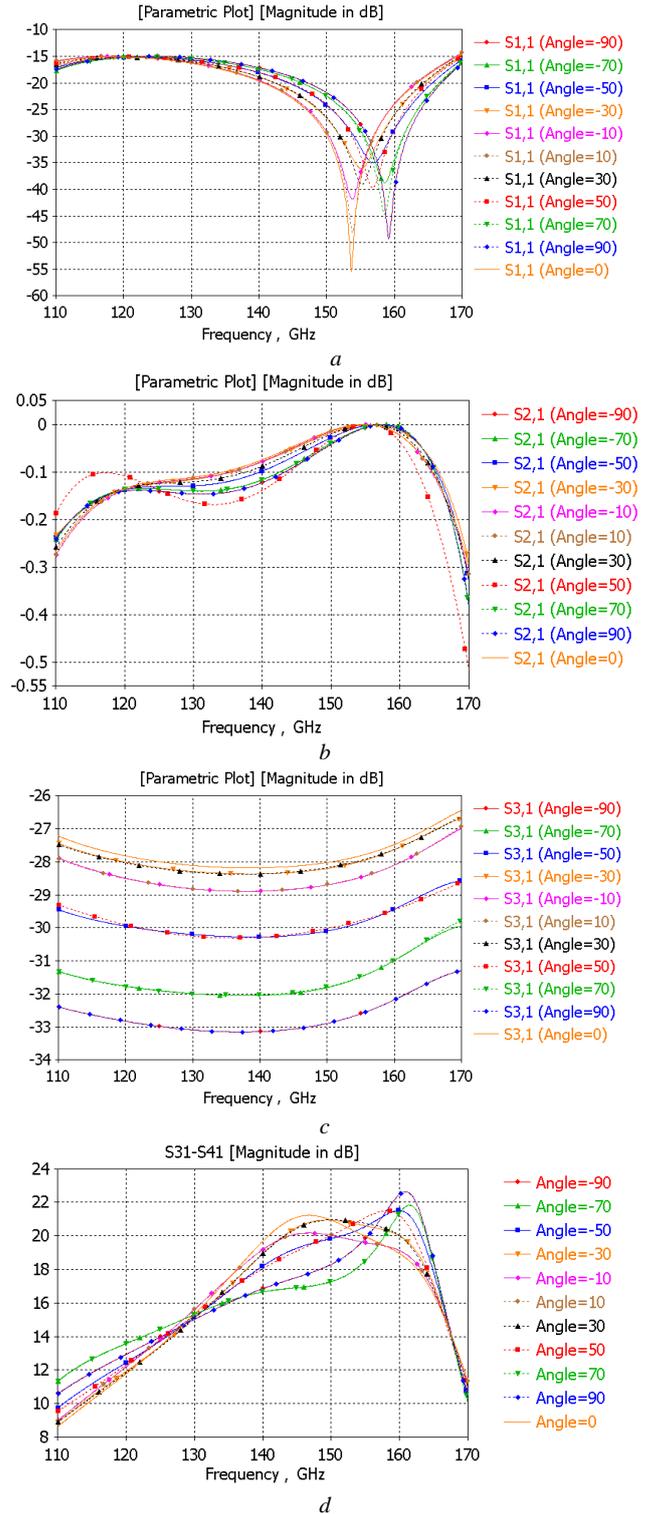

**Fig. 2** *Simulation results of the designed directional coupler.*
*a*  Return loss S11
*b*  Insertion loss S21
*c*  Coupling S31
*d*  Directivity S31-S41.

Table 1 shows the performance comparison of the proposed work with previously reported coupler in the available literature. A novel coupling method is proposed in this letter. This design has wider bandwidth, and the coupling can be tuned by changing the angle of the coupling grid. This designed coupler is suitable for ultra-wideband and large dynamic range applications.

**Table 1:** Performance comparison among proposed and reported couplers.

| Ref. | Coupling method | Frequency (GHz) | Coupling (dB) | Directivity (dB) | Return loss (dB) |
|---|---|---|---|---|---|
| This work | Coupling grid | 110 to 170 | -28.2 to -33.2 at 140 GHz | ≥9 in D band, >16 at 140GHz | ≤-15 in D band |
| [4] | Hole coupling | 22 to 26 | -20 at 24GHz | ≥9 in 2 GHz range | ≤-15 in 3 GHz range |
| [5] | Hole coupling | 26 to 40 | -20 at 33GHz | ≥9 in 10 GHz range | ≤-15 in 14 GHz range |
| [6] | CRLH | 24.5 to 28.5 | -15 at 26.5GHz | About 5 | ≥-12 dB |

The assembly of the directional coupler for D-band millimeter-wave applications is designed too, which is shown in Fig. 3. The flange adopts UG-387/U-M standard square flange. The flange and the metal shell are made of copper-plated material.

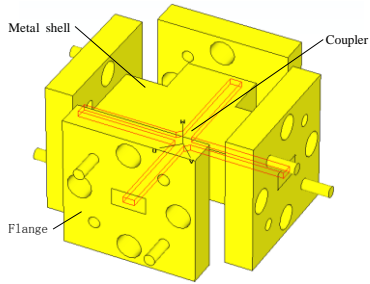

**Fig. 3** *Assembly of the designed directional coupler.*

Actually, we can also design the directional couplers used in other bands from several GHz to several hundreds of GHz based on the coupling grid proposed in this letter. The only thing we should to do is to change the size of all structures.

As shown before, the coupling can be changed by changing the angle of the coupling grid. So we design a tuning mechanism shown in Fig. 4. A putter is fixed at the edge of the metal strips, and the angle of the coupling grid can be controlled by controlling the forward or backward movement of the putter. The putter is made of dielectric materials with high hardness, such as Poly Tetra Fluoro Ethylene (PTFE). The movement of the putter can be realized manually or by using a stepper motor. When the putter moves back or forth and drives the coupling grid to rotate, it will be caused to move left or right, too. Therefore, an elongated slit needs to be formed at the position where the putter extends out of the coupler metal shell. There are shafts at the center of the two ends of each piece of metal strip, which are inserted into the holes in the metal shell to realize the rotating support of the metal strip.

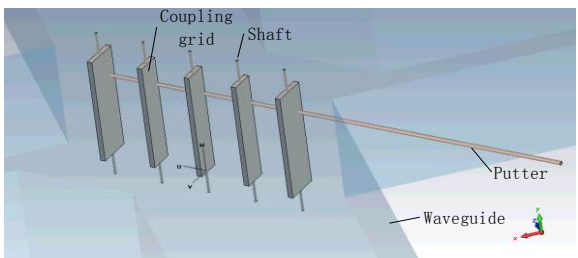

**Fig. 4** *Tuning structure of the designed directional coupler.*

*Discussion:* In the preceding section, a coupler with the coupling from -28.2 dB to -33.2 dB at 140 GHz is designed. Actually, we can design a 3-dB coupler use the similar coupling structure, which is shown in Fig. 5. The angle of the coupling grid is set to -60°. All metal strips have the same thickness, which is 0.02 mm.

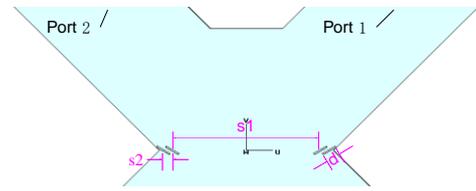

**Fig. 5** *Structure of the designed 3-dB directional coupler. Where s1=1.984 mm, s2=0.12 mm, d=0.1 mm.*

The simulation results of the designed 3-dB coupler is shown in Fig. 6. At the center frequency of 140 GHz, S21=S31=3.3144 dB. Actually, the center frequency can be changed by changing the distance of s1. This design has wider bandwidth than that reported previously in the available literature [7, 8].

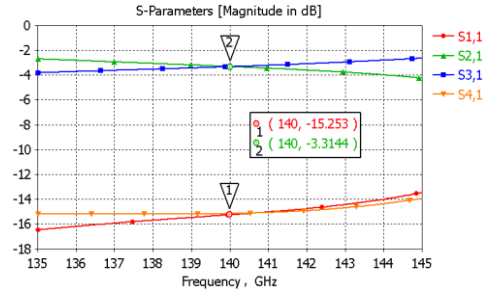

**Fig. 6** *Simulation results of the designed 3-dB directional coupler.*

*Conclusion:* A coupling tunable D-band directional coupler is designed based on a novel coupling grid structure proposed in this letter. The designed directional coupler has excellent performance with wider bandwidth. It is suitable for ultra-wideband or large dynamic range applications. A 3-dB coupler use the similar coupling structure is also designed. The coupling is 3.3144 dB at the center frequency of 140 GHz. Actually, the directional couplers used in other bands from several GHz to several hundreds of GHz can also be designed using the coupling grid structure proposed in this letter.

*Acknowledgments:* This work was supported in part by the National Magnetic Confinement Fusion Science Program of China (Grant 2015GB103000 and Grant 2015GB102003).

Weiye Xu, Handong Xu and Fukun Liu (*Institute of Plasma Physics, Chinese Academy of Sciences, Hefei, Anhui, 230031, China.*)

E-mail: xuweiye@ipp.cas.cn